# Making Graphene Nano Inductor Using Table Top Laser Engraver

Benjamin Barnes[a,b], Ibrahim Elkholy[c], Nathan Bane[c], Justin Derickson[c], Kausik S Das[a,*]

[a]Department of Natural Sciences, University of Maryland Eastern Shore, Princess Anne, MD
[b]Department of Chemistry and Biochemistry, University of Maryland College Park, MD 20742 USA
[c]Department of Engineering, University of Maryland Eastern Shore, 1, Backbone Road, Princess Anne, MD 21853 USA

**Abstract**

There is great interest in so-called nano-electronic devices due to the furious rate of device miniaturization. Fabrication of micro and nano scale resistors and capacitors have already been achieved steadily, but so far, there has been little development in the way of nano-scale coil inductors. This is because of the physical limitations in miniaturization of the design of a solenoid with wires coiled around a metallic core. So, while transistors get steadily smaller, basic inductors in electronics remained relatively bulky. Few methods exist for creating conductive polymer coils and graphene-based kinetic nano-inductors, but their large-scale fabrication process is complex and mostly beyond the current commercial technology available. So, a simpler, scalable, and robust fabrication technique is needed to overcome this bottleneck. In this work we demonstrate a new technique consisting of the laser lithography using a laser engraver of a (poly)vinyl alcohol (PVA)/graphene oxide film composite which results in a large inductive effect. We attribute this behavior to the formation of high curvature twisted screw dislocation type conductive pathways composed of polyacetylene chains linked by pi-pi interactions to reduced graphene oxide flakes resulting in inductive effect.

*Keywords:* Graphene oxide/(poly)vinyl alcohol composite, inductance, laser lithography,conductive polyacetylene/graphene Riemann surface.

## 1. Introduction

The inductor is one of four fundamental circuit components and is designed to store electrical energy in a magnetic field. Historically, inductors have been constructed by physically coiling wires around a paper, plastic, or metal core [1], but more recently, advances in printed circuit board (PCB) fabrication and photolithographic techniques have led to the development of planar inductors consisting of a tiny conductive spiral etched or deposited onto a substrate [2, 3]. While these systems exploit the latest microelectronic fabrication techniques, they are not without problems such as a generally low-quality factor, high ohmic losses, as well as a relatively large area footprint, so alternatives are being sought in the field of molecular electronics[4]. Recent modeling has proposed that certain spiral configurations of graphene formed by screw dislocations can form Riemann surfaces capable of acting as nanoinductors[4] but there exist no means of fabricating or isolating such structures for use in circuits. There are also methods for synthesizing coiled polymer structures by using a liquid crystal substrate, but the electrical properties of these materials have not been probed extensively and the synthetic technique is too complex for industrial production. In order to maintain the pace of miniaturization, a simple scalable means of fabricating nanoscale inductor coils is needed.

---

*Corresponding author
*Email address:* kdas@umes.edu (Kausik S Das)

In this work we demonstrate such a method of fabricating inductive carbon nano-twists (ICTs) embedded within laser lithographed PVA/GO films Fig. 1. FTIR analysis is used to determine the nature of the materials present in the lithographed films (Fig. 4) Impedance spectroscopy and direct LCR circuit resonance and phase response are used to characterize the inductive properties of films with variable composition, thickness, and macrostructure in order to determine the structure and distribution of the ICTs in the films.

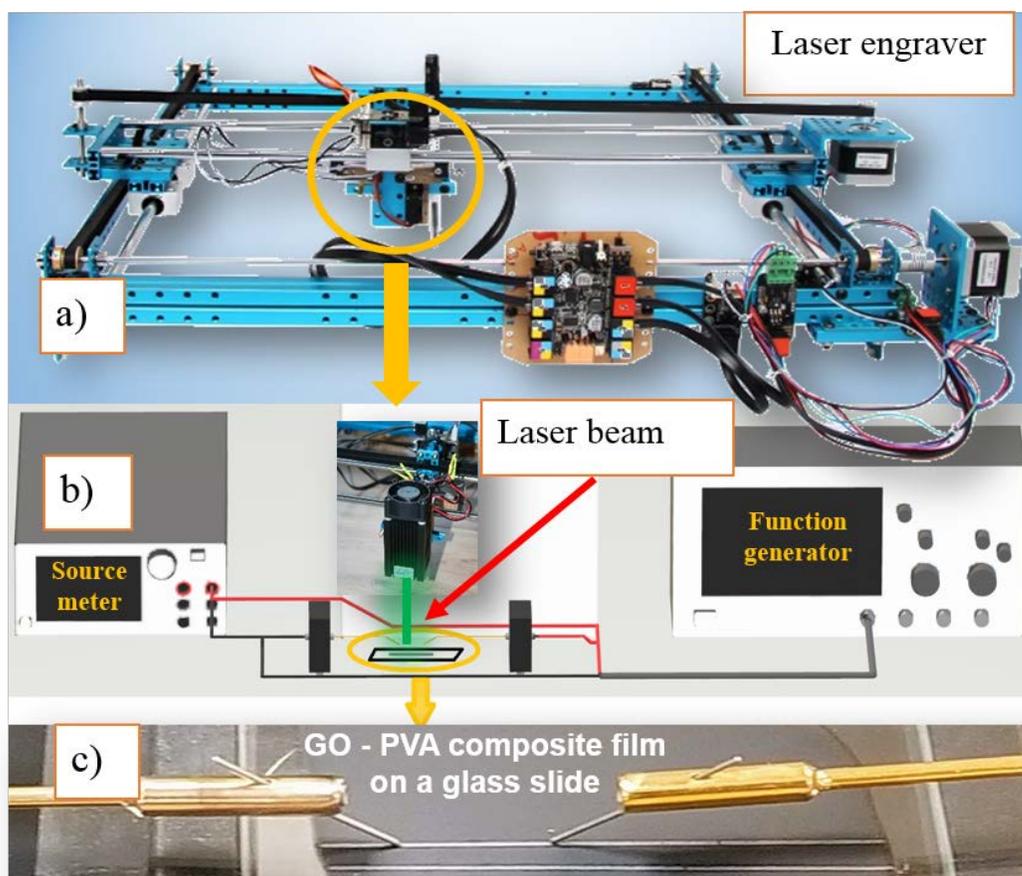

Figure 1: Schematic of laser scribed graphene nano inductor formation. a) A commercial x-y plotter/laser engraver is used to scribe graphene oxide and convert it to graphene. b) A graphene line is created by scribing a GO/PVA composite thin film by laser. The reduced graphene inductor is used to construct a LCR circuit to investigate resonance behavior. c) A close-up view of the laser scribed graphene inductor with micro-manupulator connectors.

## 2. Results

## 3. Experimental Methods

PVA/GO composites were made by combining PVA powder with water at 90 °C, stirring until a clear solution was attained, and then diluting the resulting gel in aqueous GO solution to obtain a range of final PVA:GO mass ratios in aqueous solution. These were drop-casted on PET films and heated in air at 50 °C until the water had evaporated. Unless otherwise noted, the resulting films were on the order of 100 $\mu$m thick. PVA/GO thin films thus prepared were then subjected to laser lithography (Fig. 1a-c) using a commercial laser engraver/cutter (405 nm). A number of recent papers have propelled the method of fabricating carbon electrodes from laser-reduced graphene oxide (GO) films into the realm of commercial feasibility[5, 6, 7], but none have investigated the techniques impact on GO/PVA composite films. This method is an attractive means of GO reduction as it serves to simultaneously exfoliate and reduce the GO resulting in the liberation of oxygen as CO, $CO_2$, and $H_2O$ gas[5, 6]. By tuning parameters such as laser intensity, scribing speed, and substrate temperature, the conductivity and overall quality of the laser scribed graphene (LSG) layer can be precisely controlled[6, 8, 7, 9]. An additional benefit of this fabrication technique is the high level of structure and detail attainable thanks to the precision of modern laser lithography techniques and instruments[10].

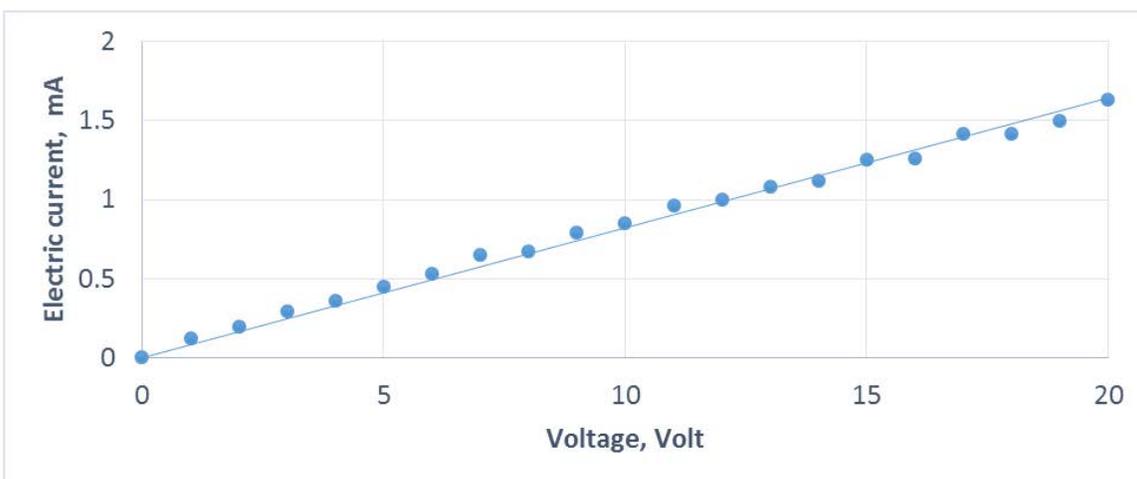

Figure 2: Ohmic behavior of the laser scribed graphene line.

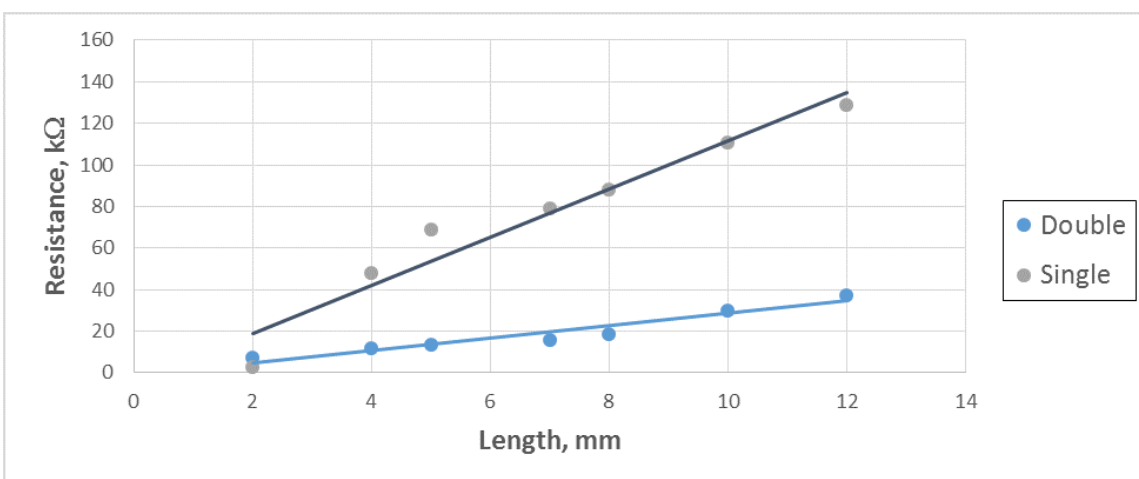

Figure 3: Change of resistance as a function of the length of the graphene line.

For an accurate and reproducible analysis of the electronic properties of the films, copper leads were attached to the desired position in the scribed area with carbon glue. Alternatively, micromanipulator probes were used to characterize the electrical response (Fig. 1c). The leads were on the order of 1 cm long to prevent lead effects from impacting the impedance analysis.

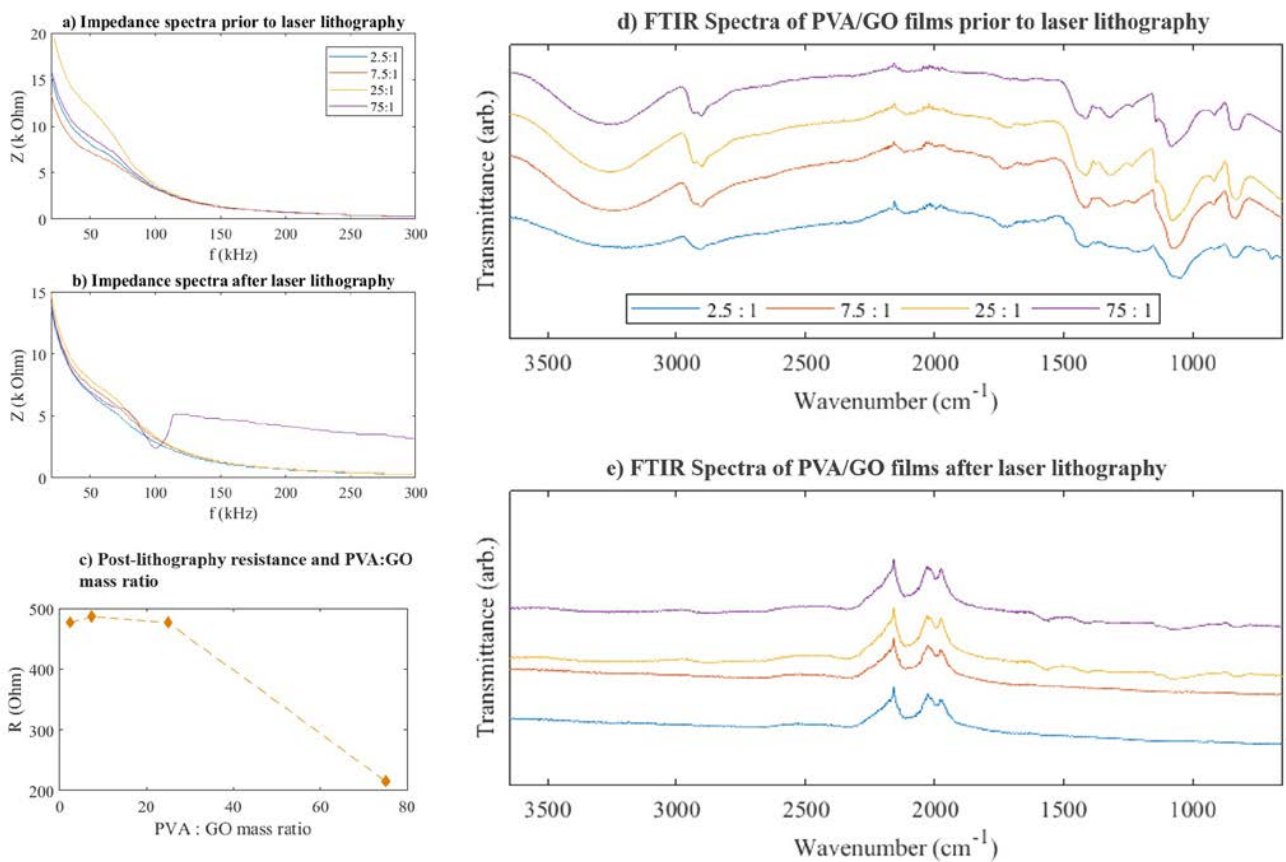

Figure 4: Changes in the electrical properties of PVA/GO films with varying PVA:GO ratios. A) The impedance characteristics of non-lithographed PVA/GO films is similar across all PVA:GO.ratios. A capacitive regime is demonstrated at low frequencies (¡150 kHz) followed by a more gradual decay in the slope. B) After laser lithography, only the film with the highest PVA:GO ratio demonstrated a change in impedance characteristics, with the development of a self-resonant frequency peak at ∼ 100 kHz followed by a decline in the slope at greater magnitudes of inductance. C) The resistance of lithographed films was least for that with the highest PVA:GO ratio but remained roughly constant for low PVA:GO ratios. D) The FTIR spectra of films of all compositions was approximately the same prior to laser lithography. E) The FTIR following laser lithography demonstrate the loss of all OH and CO bonds, confirming the reduction of GO to graphene. Also of interest is the development of peaks for a CC conjugated system in the films with the highest PVA:GO ratios, indicating the formation of polyacetylene from the decomposition of PVA.

Linear Ohmic behavior of the current-voltage characteristics of a laser scribed lines is shown in Fig. 2, whereas the variation of resistance of the scribed lines as a function of length is shown in Fig. 3.

Changes in electrical properties of PVA/GO films with varying ratios are shown in Fig. 4. The resistance of the laser scribed films of varying PVA:GO ratios was measured at an applied potential of 2 V and plotted in Fig. 4c. These data illustrate that at a threshold PVA concentration, the resistance of scribed films begins to decrease. At low PVA:GO ratios, the resistance remained high at around 0.5 kΩ which dropped to 0.2 kΩ at a PVA:GO ratio of 75:1. $^{13}$CNMR and CP/MS demonstrated in previous literature[11] that at moderately high temperatures, such as those generated locally with the laser lithography process, PVA decomposes to polyacetylene (PA). The presence of polyacetylene in the lithographed PVA/GO composites of higher PVA concentration is supported in our work by the small band at 1550 $cm^{-1}$ in the FTIR spectrum indicating a conjugated system of carbon double bonds (Fig. 4e). Polyacetylene in its natural state is a semiconductor, and only becomes appreciably conductive at standard conditions when doped. It seems then, that in these composites, PA is becoming doped by the presence of either the LSG or the gaseous by-products of the reduction process thus improving the overall conductivity of the lithographed composites.

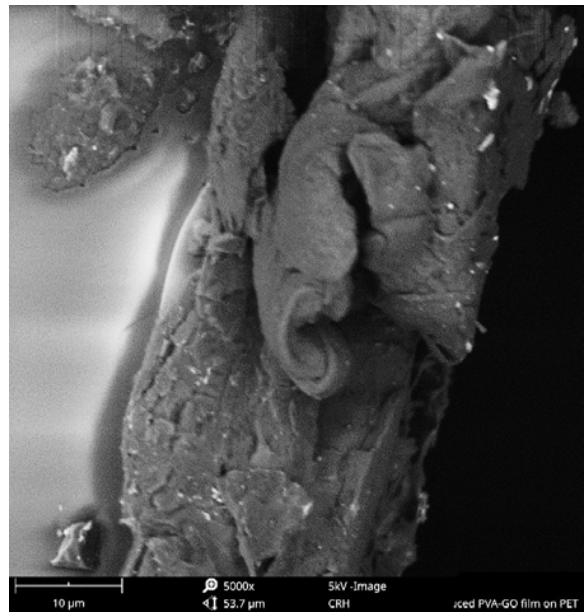

Figure 5: Cross-sectional view of laser scribed graphene lines showing twisting of conductive reduced graphene oxide pathways.

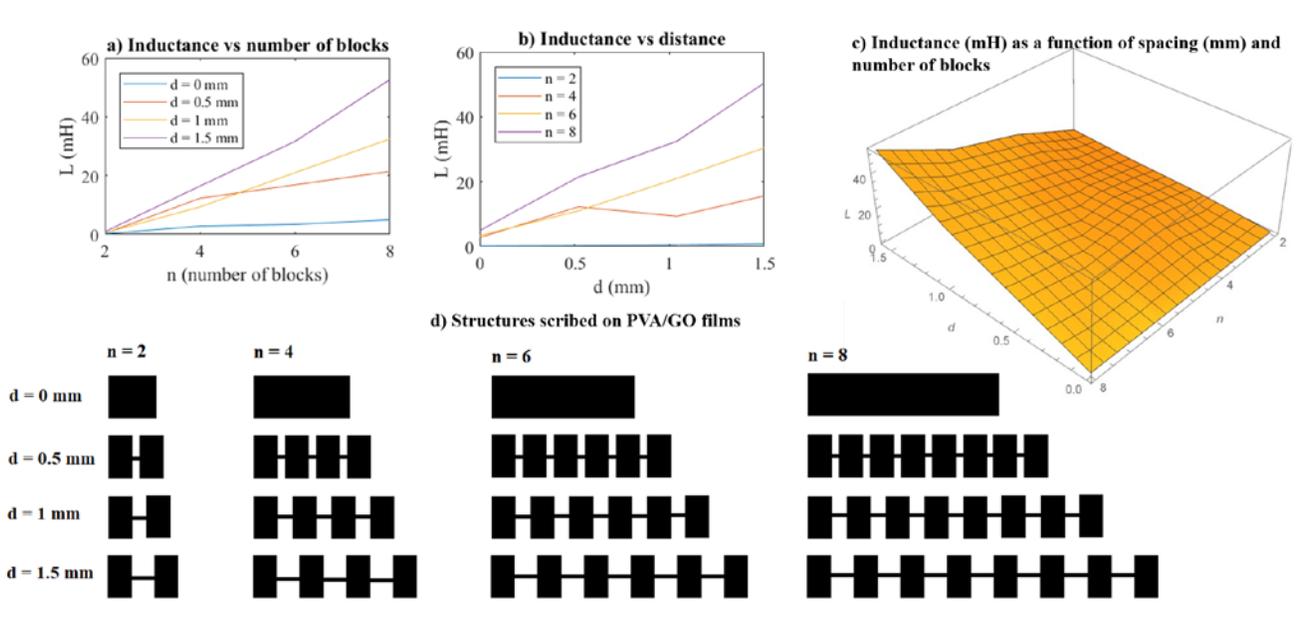

Figure 6: The growth of inductance with respect to surface area (number of blocks) distancing of blocks (distance). A) Inductance was shown to grow nearly linearly with surface area for all spacings. However, the rate of growth increased with distance. The average rate of growth of inductance with surface area is ∼ 3.7. B) The growth of inductance with distance is nearly linear for surface areas of 2, 6, and 8. The average rate of growth with distance is ∼ 17, much larger than that for surface area. C) A summary of both trends with a single area-distance surface. D) the structures used to analyze the impact of surface are and spacing on inductance.

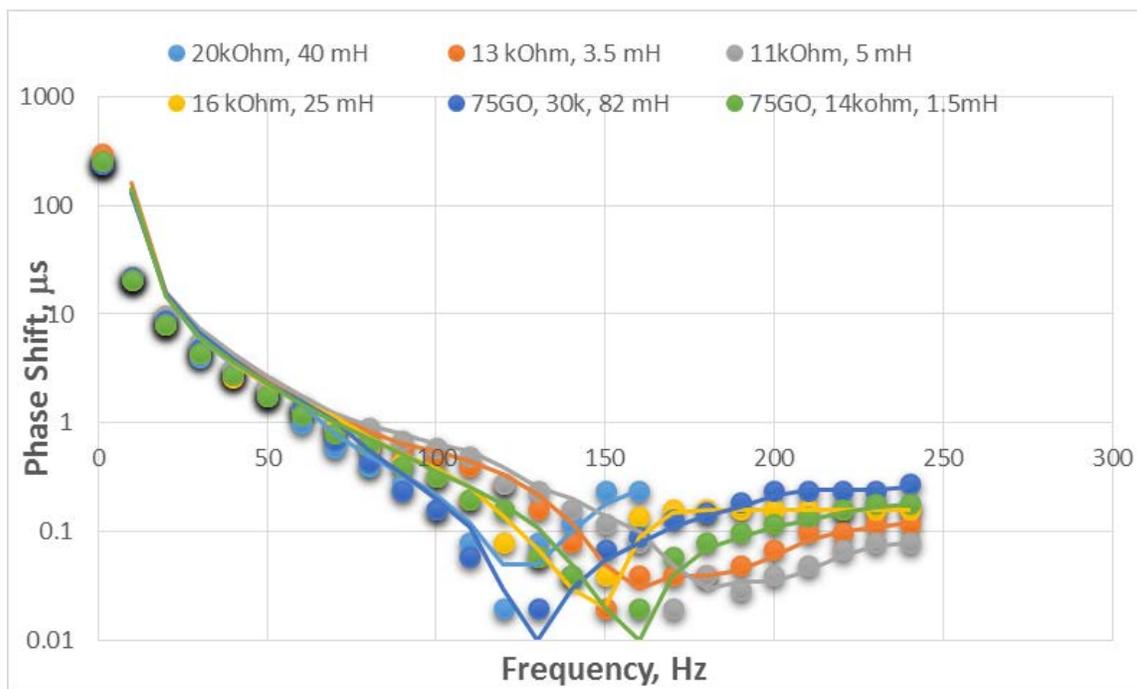

Figure 7: Changes in phase shift for a 75% GO sample. A series LRC circuit was constructed with the laser scribed graphene line acting as an inductor. An external capacitor (C=40 pf) and an external resistance (R=67 kΩ) is connected in series. An input function generator was used to vary the input frequency and the output phase difference between the input and the output across the graphene line is measured. It clearly shows resonance indicating the inductive nature of the graphene line. Frequency dependent impedance is also noted for different input frequencies.

Impedance analysis of these films shows interesting behavior: At all PVA:GO ratios, the impedance spectra were essentially the same prior to laser lithography (Fig. 4a), with a capacitive regime up to 150 kHz test frequency, followed by a flat regime which persisted across the remaining test frequencies (300 kHz). At low PVA:GO ratios (2.5, 7.5, and 25), the behavior remained the same following laser lithography (Fig. 4a), but for the highest PVA:GO ratio, a clear self-resonant frequency area becomes visible at 100 kHz, followed by a region of higher impedance up to 300 kHz, indicating the start of more dominant inductive behavior at high PVA:GO ratios.

Thus far, only simple devices were tested consisting of a single laser scribed rectangular area. Next, the effect of device macrostructure was evaluated at a uniform PVA:GO ratio and uniform thickness, but with different patterns consisting of uniform block structures as illustrated below in Fig. 6. The blocks were sorted by the number of blocks in each circuit (n) and spacing between blocks (d). The resistance was measured for each structure by placing test probes at opposite ends of a block or row of blocks. This demonstrated linear trends in both resistance vs spacing between blocks and resistance vs number of blocks.

The impedance spectra of the different structures showed similar behavior for all. The maximum inductance occurred at different frequencies for the structures, yielding no discernable frequency trend, but the magnitude of the maximum inductance demonstrated a linear correlation with the number of blocks in each structure and again with the distance between the blocks. These values are recorded in Fig. 6a and Fig.6b. As shown in Fig. 4a, the maximum inductance was small and increased only slightly as the size of the blocks increased (d = 0, n = 2, 4, 6), but grew much more dramatically with the spacing showing that the inductive effect is augmented by the spacing of the structures more so than by the surface area (Fig. 6b). An investigation of phase behavior and resonance with respect to the input frequency is shown in Fig.7 and corresponding inductance as a function of the length of the scribed line is shown in Fig.8. Changes of resistance with different lengths of the laser scribed inductors as a function of the resonant frequency for different GO samples can be seen in Fig.9.

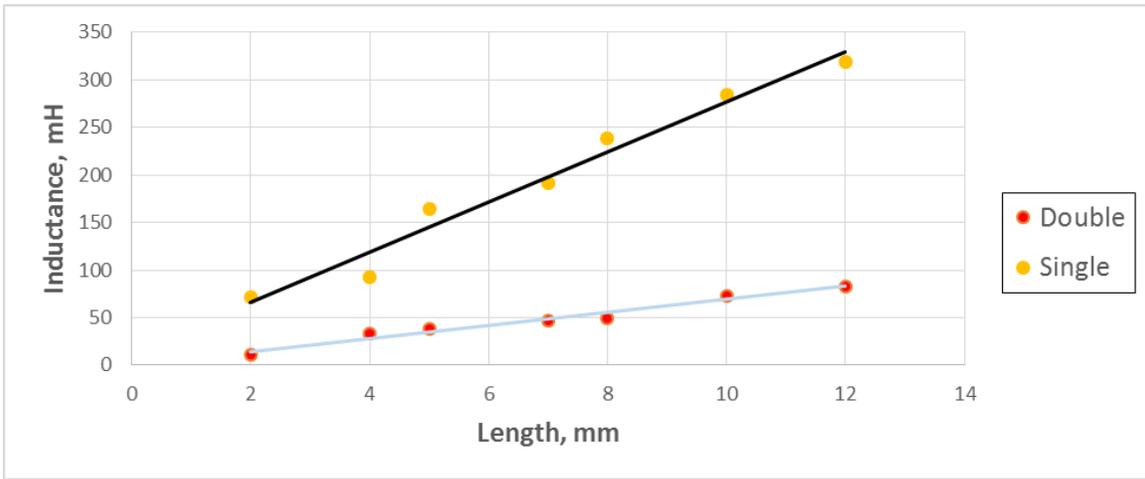

Figure 8: Change of inductance of the graphene line as a function length and number of rastering.

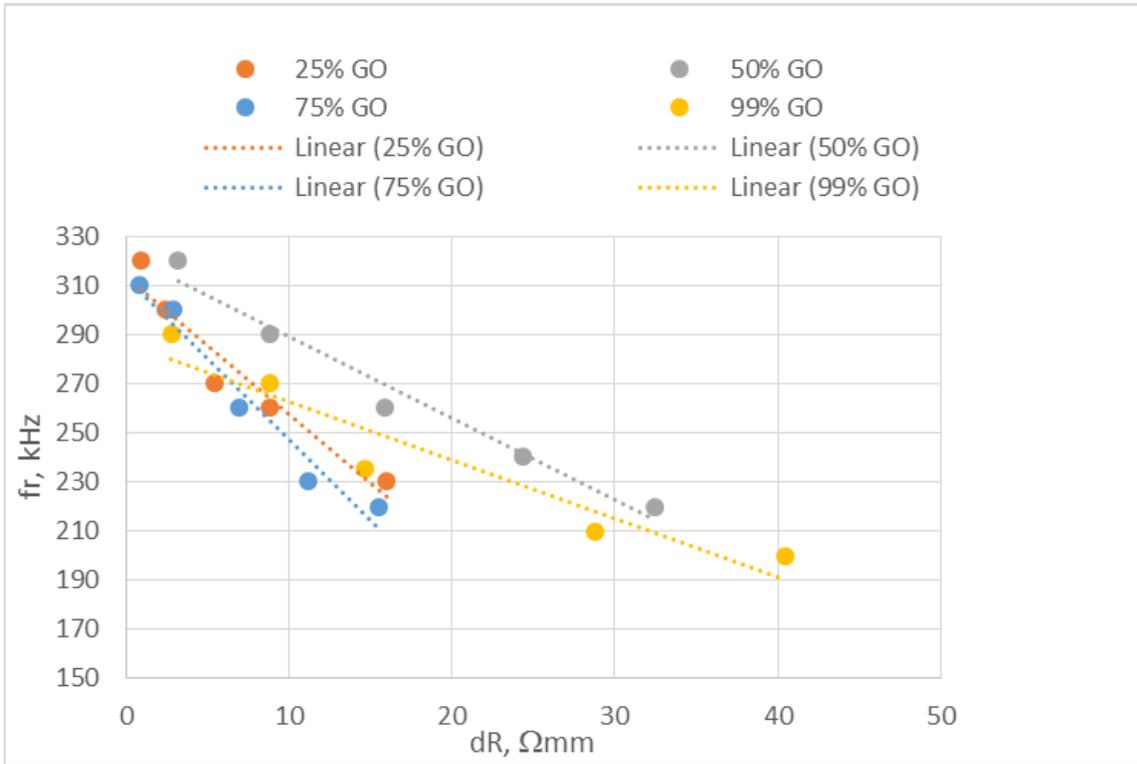

Figure 9: Changes of resistance with different lengths of the laser scribed inductors as a function of the resonant frequency for different GO samples.

In this experiment, the effect of distance and the number of blocks (surface area) can easily be decoupled by analyzing the average rate of growth of inductance with respect to each parameter (Fig. 6c). Such an analysis shows that impedance grows with distance at a rate of 17x, whereas it grows with the number of blocks by about 3.7x. This indicates that the distribution of ICNs is discrete rather than continuous throughout the PVA/GO film. Hence the number of alternative channels through which the current can flow within the blocks increases with surface area, thus allowing the electrons to circumvent most of the ICNs; however, when blocks are spaced and connected by a thin lithographed line, the inductive effect is observed most strongly, as the electrons, when travelling across the lines, have fewer alternative channels available and hence must pass through the ICNs located on the respective channels. This hypothesis was tested by the development of an experiment at a PVA:GO ratio of 75:1, constant surface area consisting of a single laser lithographed region, but with variable film thickness ( 25 130 $\mu$m). If there is indeed a distribution of ICNs throughout the film, one would expect a window of film thicknesses to exhibit significant inductive behavior, as the thinnest films may not be thick enough to host an ICN, but the thickest films will provide alternative pathways for electrons to travel through. The impedance spectrum prior to laser lithography was similar for all thicknesses (Fig. 4a), exhibiting the same behavior as before, as expected. However, after laser scribing the surface, a clear trend in the impedance spectra developed lending support to our hypothesis (Fig. 4): the thinnest film exhibited behavior as if unchanged from that of the non-laser scribed film, with a small capacitive region up to $\sim$ 100 kHz, but as the film thickness increased, inductive behavior was observed for all samples, with a strong positive slope for much of the impedance spectrum. Additionally, this inductive effect remained, but decayed in magnitude as the film thickness was increased, indicating the development of vertical channels by which electrons could avoid ICNs.

Through numerous computational and imaging studies, polyacetylene has been shown to form rich 3D structures[12]. In its isolated state, PA is a planar molecule[12], but when either grown in the presence of a chiral matrix, such as liquid crystal substrate[13], or placed in the presence of other highly structured pi-pi systems such as carbon nanotubes[14], it can assume complex macromolecular structures such as helices[15, 16], coils, and supercoils. While our system lacks known chiral surfaces and tubular structures, it does contain a dispersion of graphite oxide flakes, which, as discussed before, when irradiated with laser light, reduce to form graphene. We propose that the presence of graphene stacks directs the shape-evolution of the PA structure through pi-pi interactions as it is formed from the decomposition of PVA chains. This propagation of a PA/graphene complex through the film appears to provide enough structure to form a dispersion of crude nano-scale coil inductors throughout a given PVA/GO film following laser lithography.

## 4. Conclusion

In this work we have shown that subject to pulsed laser scribing PVA/GO composite using a table top commercial laser cutter/engraver exhibits inductive effects in reduced GO lines. We hypothesize that as laser scans through the PVA-GO film, it periodically heats up the composite. Under laser irradiation dielectric GO transforms into conductive reduced GO. Furthermore, we hypothesize that periodic heating along the scanning line tends to twist the graphene flakes, thereby creating an equivalent Riemann surface and a source of inductance. This simple fabrication technique may pave the way for graphene micro/nano inductors. A cross sectional SEM image at the edge of the laser scribed line indeed show twisting of graphene flakes embedded in PVA (Fig.5) creating an equivalent conducting nano coils capable of producing inductive effects. A comparison of inductance as a function of length of the graphene lines is shown in Fig. 8.

## 5. Acknowledgement

This work was partially supported by the National Science Foundation (Award # 1719425), the Department of Education (Award # P120A70068) and Maryland Technology Enterprise Institute through MIPS grant.

**References**

[1] E. M. Purcell, D. J. Morin, Electricity and magnetism, Cambridge University Press, 2013.

[2] W. Ni, J. Kim, E. C. Kan, Permalloy patterning effects on rf inductors, IEEE Transactions on Magnetics 42 (2006) 2827–2829.

[3] S. Liu, L. Zhu, F. Allibert, I. Radu, X. Zhu, Y. Lu, Physical models of planar spiral inductor integrated on the high-resistivity and trap-rich silicon-on-insulator substrates, IEEE Transactions on Electron Devices 64 (2017) 2775–2781.

[4] F. Xu, H. Yu, A. Sadrzadeh, B. I. Yakobson, Riemann surfaces of carbon as graphene nanosolenoids, Nano letters 16 (2016) 34–39.

[5] R. Trusovas, K. Ratautas, G. Račiukaitis, J. Barkauskas, I. Stankevičienė, G. Niaura, R. Mažeikienė, Reduction of graphite oxide to graphene with laser irradiation, Carbon 52 (2013) 574–582.

[6] V. Strong, S. Dubin, M. F. El-Kady, A. Lech, Y. Wang, B. H. Weiller, R. B. Kaner, Patterning and electronic tuning of laser scribed graphene for flexible all-carbon devices, ACS Nano 6 (2012) 1395–1403.

[7] M. F. El-Kady, R. B. Kaner, Scalable fabrication of high-power graphene micro-supercapacitors for flexible and on-chip energy storage, Nature communications 4 (2013) ncomms2446.

[8] Y. C. Guan, Y. Fang, G. Lim, H. Zheng, M. H. Hong, Fabrication of laser-reduced graphene oxide in liquid nitrogen environment, Scientific Reports 6 (2016) 28913.

[9] S. Evlashin, P. Dyakonov, R. Khmelnitsky, S. Dagesyan, A. Klokov, A. Sharkov, P. Timashev, S. Minaeva, K. Maslakov, S. Svyakhovskiy, et al., Controllable laser reduction of graphene oxide films for photoelectronic applications, ACS applied materials & interfaces 8 (2016) 28880–28887.

[10] A. Pendurthi, S. Movafaghi, W. Wang, S. Shadman, A. P. Yalin, A. K. Kota, Fabrication of nanostructured omniphobic and superomniphobic surfaces with inexpensive $co_2$ laser engraver, ACS Applied Materials & Interfaces 9 (2017) 25656–25661.

[11] J. W. Gilman, D. L. VanderHart, T. Kashiwagi, Thermal decomposition chemistry of poly (vinyl alcohol) char characterization and reactions with bismaleimides, ACS Publications, 1995.

[12] K. Akagi, T. Mori, Helical polyacetyleneorigins and synthesis, The Chemical Record 8 (2008) 395–406.

[13] M. Goh, S. Matsushita, K. Akagi, From helical polyacetylene to helical graphite: synthesis in the chiral nematic liquid crystal field and morphology-retaining carbonisation, Chemical Society Reviews 39 (2010) 2466–2476.

[14] M. Shan, Q. Xue, T. Lei, W. Xing, Z. Yan, Self-assembly of helical polyacetylene nanostructures on carbon nanotubes, The Journal of Physical Chemistry C 117 (2013) 16248–16255.

[15] Y. Li, H. Fu, W. Lu, S. Xu, X. Zhang, Infinite spinning of several polyacetylene chains into long multiple helices, Carbon 123 (2017) 62–69.

[16] T. Miyagawa, A. Furuko, K. Maeda, H. Katagiri, Y. Furusho, E. Yashima, Dual memory of enantiomeric helices in a polyacetylene induced by a single enantiomer, Journal of the American Chemical Society 127 (2005) 5018–5019.